# Observation of fractional topological numbers at photonic edges and corners

Chengpeng Liang[#]

*School of Electronic Science and Engineering, Nanjing University, 163 Xianlin Road, Nanjing, 210023, China*

Yang Liu[#]

*Institute of Theoretical and Applied Physics, School of Physical Science and Technology & Collaborative Innovation Center of Suzhou Nano Science and Technology, Soochow University, 1 Shizi Street, Suzhou, 215006, China*

Fei-Fei Li[#]

*School of Electronic Science and Engineering, Nanjing University, 163 Xianlin Road, Nanjing, 210023, China*

Shuwai Leung[#]

*School of Electronic Science and Engineering, Nanjing University, 163 Xianlin Road, Nanjing, 210023, China*

Yin Poo[†]

*School of Electronic Science and Engineering, Nanjing University, 163 Xianlin Road, Nanjing, 210023, China*

Email: ypoo@nju.edu.cn

Jian-Hua Jiang[†]

*Institute of Theoretical and Applied Physics, School of Physical Science and Technology & Collaborative Innovation Center of Suzhou Nano Science and Technology, Soochow University, 1 Shizi Street, Suzhou, 215006, China*

*Key Lab of Advanced Optical Manufacturing Technologies of Jiangsu Province & Key Lab of Modern Optical Technologies of Ministry of Education, Soochow University, Suzhou, 215006, China.*

Email: jianhuajiang@suda.edu.cn

[#] These authors contributed equally to this work.

[†] To whom correspondence should be addressed.

## Abstract

**Topological phases of matter are featured with exotic edge states. However, the fractional topological numbers at edges, though predicted long ago by Jackiw and Rebbi, remain elusive in topological photonic systems. Here, we report on the observation of fractional topological numbers at the topological edges and corners in one- and two-dimensional photonic crystals. The fractional topological numbers are determined via the measurements of the photonic local density-of-states. In one-dimensional photonic crystals, we witness a rapid change of the fractional topological number at the edges rising from 0 to 1/2 when the photonic band gap experiences a topological transition, confirming the well-known prediction of Jackiw and Rebbi. In two-dimensional systems, we discover that the fractional topological number in the corner region varies from 0 to 1/2 and 1/4 in different photonic band gap phases. Our study paves the way toward topological manipulation of fractional quantum numbers in photonics.**

## Introduction

Early in 1976, Jackiw and Rebbi predicted from quantum field theory the emergence of a topological boundary state and a fractional quantum charge $e/2$ at the mass domain-wall of one-dimensional (1D) massive Dirac fermions [1] (Fig. 1a). This prediction sets the foundation for understanding the topological edge states in the Su-Schrieffer-Heeger (SSH) model [2] (Fig. 1b) and later in various other topological materials and systems [3, 4]. In the past forty years, topological edge states served as a main experimental signature of topological phases of matter, except in a few experiments [5-8].

At the heart of topological photonics [9-17], there are topological edge and corner states as well as the fractional topological numbers (i.e., the photonic analog of fractional charges) at these boundaries. Many efforts have been devoted to the study of topological boundary states. For light-matter interacting systems [18, 19] (such as lasing [20] and quantum electrodynamics [21, 22]), the fractional topological numbers which are akin to the photonic local density-of-states (LDOS) also play a key role. Besides, fractional topological numbers serve as a more robust signature of topological phases [23-29], even when the chiral symmetry is broken as often the case in photonic systems. However, to date, these edge and corner fractional topological numbers have not yet been confirmed in photonic experiments.

Here, we report on the experimental observation of fractional topological numbers at photonic topological edges and corners. Exploiting the SSH physics, we design a series of 1D and two-dimensional (2D) dielectric photonic crystals (PCs) that support fractional topological numbers at the edge and corner boundaries. The SSH model, initially proposed in 1979 [2], predicts the emergence of a topologically localized state and the fractional charge $e/2$ at each soliton in polyacetylene. Due to electron spin degeneracy and other technical challenges, such a fractional charge, however, has not yet been observed in polyacetylene or any other condensed matter systems. Here, using the nonequilibrium nature of photons (i.e., photons can be excited at any frequency to probe the LDOS), we show that fractional topological numbers---which emerge due to the equilibrium filling of electrons according to the LDOS---can instead be revealed in photonic systems. We confirm experimentally the emergence of the half quantum number in the 1D SSH model using dielectric PCs. We witness a rapid change of the fractional topological number across a topological transition of the 1D photonic system by tuning the geometry of the PCs. We further reveal in experiments the 1/2 and 1/4 fractional topological numbers at the edges and corners of various 2D photonic topological insulators.

## Results

In the 1D SSH model, the Wannier center, which represents the center-of-mass location of the electrons in the valence band, can be tuned by the nearest-neighbor couplings. When the intra-cell coupling $t_0$ is stronger (weaker) than the inter-cell coupling $t$, the Wannier center is at the unit-cell center (edge) [23-25], as labeled by the stars in Fig. 1c. The former (latter) case is regarded as trivial (topological) and denoted as phase $\alpha$ ($\beta$). Since electrons' average position determines the bulk polarization in an insulator, the case with the Wannier center at the unit-cell center (edge) corresponds to a bulk polarization $P = 0$ ($\frac{1}{2}$). These results can also be verified via the numerical Wilson-loop approach [30].

The emergence of the fractional charge $\pm\frac{e}{2}$ at topological edges can be understood via the filling anomaly---a phenomenon that the number of valence band eigenstates is not equal to the number of unit-cells in finite systems. For instance, Fig. 1d gives the spectrum of a finite SSH chain with six unit-cells for both phases $\alpha$ and $\beta$. In phase $\alpha$ the valence band has six eigenstates, whereas in phase $\beta$ there are only five eigenstates. In phase $\beta$, when only the valence band is filled, there is an electron missing in the system. The system thus has in total a charge of $-e$ ($e$ is the charge of an electron). Because the system is inversion symmetric, there is $q_e = \frac{1}{2}|e|$ charge for half of the system (Fig. 1e). When the two edge states are also filled, the total charge of the half-sector becomes $q_e = \frac{1}{2}e$, since each localized edge state contributes charge $e$ to a half-sector. In comparison, for phase $\alpha$, the system has charge neutrality whenever the Fermi energy is in the band gap.

The fractional charge can be counted straightforwardly when the system is tuned adiabatically to the limit with $t_0 = 0$. In this limit, valence band electrons are localized at the Wannier centers. Since the Wannier center is at the unit-cell edge, the valence band filling contributes $\frac{e}{2}$ to each unit-cell from one Wannier center. As shown in Fig. 1e, all unit-cells except the edge unit-cell has charge $e$ due to the valence band filling which compensates the background positive ion charge. However, the edge unit-cell has only $\frac{e}{2}$ charge coming from the valence band filling. Counting the background positive ion charge, the edge unit-cell has a charge of $q_e = \frac{1}{2}|e|$, yielding the same results as in the above paragraph. We denote the fractional topological number at a topological edge as $Q_e = \frac{1}{2}$ which characterizes the fractional prefactor. The fractional charge also emerges at the edge boundaries between two SSH models in topologically distinct phases (see Supplementary Note 1). In fact, such edge boundaries resemble the Dirac mass domain wall in the Jackiw-Rebbi theory [3, 4], since the two phases $\alpha$ and $\beta$ can be described by 1D massive Dirac equations of opposite Dirac masses [31].

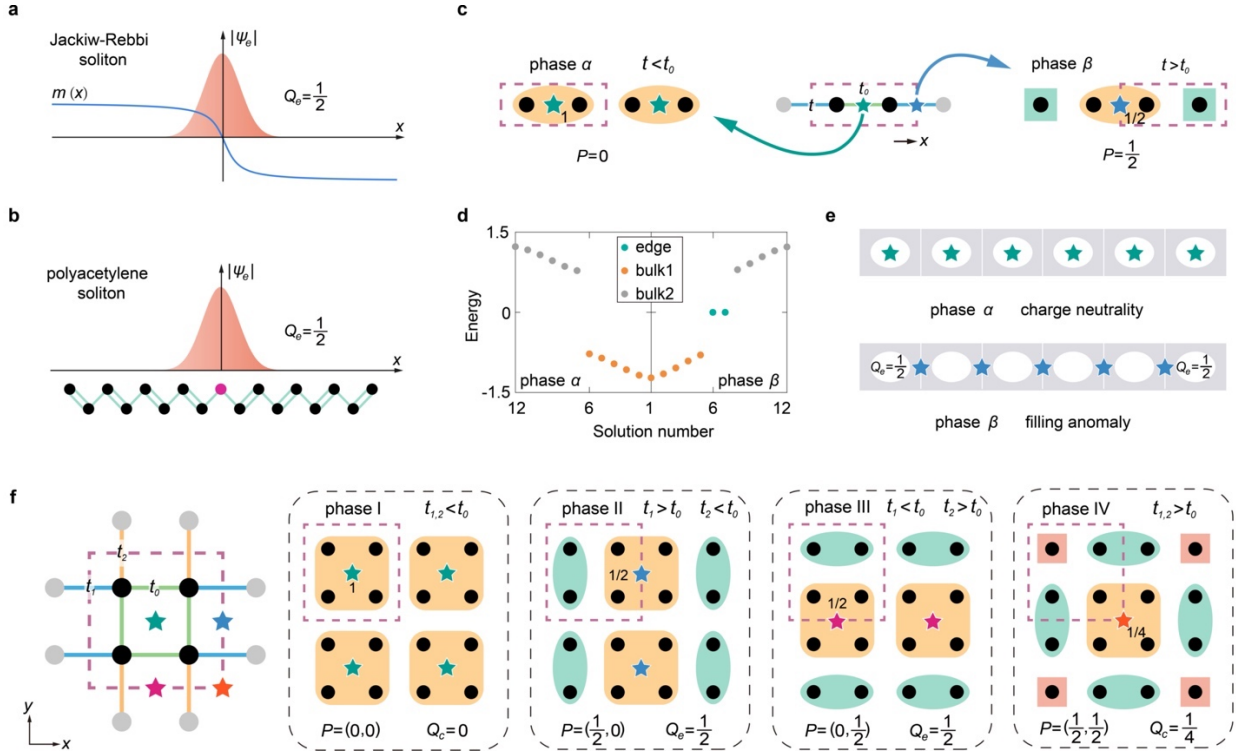

**Figure 1 | Emergent fractional charges in 1D and 2D systems. a**, A half-charge emerging at a Dirac mass domain wall in the 1D Jackiw-Rebbi model. The blue line denotes the position-dependent Dirac mass. **b**, A half-charge accompanying a soliton mode in polyacetylene (illustrated schematically in the below the $x$ axis). In **a** and **b**, the red shadows represent the wavefunctions of the topologically localized wavefunctions $|\psi_e|$. **c**, Topologically distinct phases, $\alpha$ and $\beta$, in the 1D SSH model. The unit-cells are depicted by the brown dashed boxes. The yellow regions denote the bulk modes, while the green regions denote the edge modes. **d**, Eigen-spectra for phases $\alpha$ and $\beta$ for finite chains with six unit-cells. **e**, Illustration of charge neutrality for phase $\alpha$ and the filling anomaly for phase $\beta$. **f**, Illustration of the 2D SSH model and its four different phases. In all models, dots label the lattice sites, lines connecting them label their mutual couplings, stars label the Wannier centers, and the lattice constants are set to be unity.

In the 2D SSH model [32, 33], the above scenario is enriched. There are four different cases (denoted as phases I-IV), as illustrated in Fig. 1f. When the intra-cell coupling $t_0$ is weaker (stronger) than the inter-cell coupling in the $x$ direction $t_1$, the Wannier center is shifted by 1/2 (0) in the $x$ direction. Similar scenario happens for the $y$ direction. The shift of the Wannier center is characterized by the bulk polarization vector, $\boldsymbol{P}$, which is a vector pointing from the unit-cell center to the Wannier center (see Supplementary Note 2). The phases I-IV have, respectively, $\boldsymbol{P} = (0,0), (\frac{1}{2}, 0), (0, \frac{1}{2}), (\frac{1}{2}, \frac{1}{2})$ [32, 33]. The finite polarization yields fraction quantum numbers at the edge and corner boundaries. For instance, phase II (III) has a fractional topological number for the edge boundary perpendicular to the $x$ ($y$) direction $q_e = \frac{1}{2}|e|$, (denoted by the fractional edge quantum number $Q_e = \frac{1}{2}$), due to the mechanism similar to that in the 1D SSH model. For phase IV, in addition to the fractional topological numbers for both the $x$ and $y$ edge boundaries $Q_e = \frac{1}{2}$, there is another fractional topological number emerges at the corner boundaries $Q_c = \frac{1}{4}$ due to the fact that there is a 1/4 bulk Wannier center shared by a corner unit-cell with other parts of the system [34]. This fractional topological number at corner boundary is a signature of the higher-order topology [6, 34-44].

To experimentally verify the emergent fractional topological numbers, we use dielectric PCs----fundamental materials for subwavelength photonics---to realize the SSH models. For the 1D SSH model, it is realized by PCs with two alumina pillars of identical radius $r = 0.1a$ in each unit-cell where the lattice constant $a = 2$ cm (Fig. 2a). To form 1D photonic systems, the PC is cladded by metallic walls in both the $y$ and $z$ directions (see Methods). We focus on the photonic bands formed by the fundamental modes of the 1D photonic waveguide. In each unit-cell, two pillars are placed symmetrically with respect to the unit-cell center to ensure the inversion symmetry (Fig. 2a). By tuning the distance between the unit-cell center and the center of a pillar, $d$, we can modify the photonic bands and the band topology. As shown in Figs. 2b and 2c, for large $d$, there is a parity inversion in the first photonic band between the Γ and the X points in the Brillouin zone, signaling that a topological phase (i.e., the phase $\beta$) where the Wannier center locates at the unit-cell edge [23]. In contrast, for small $d$, there is no such parity inversion. Thus, the Wannier center is at the unit-cell center and the system is trivial (i.e., the phase $\alpha$) [23]. The topological transition between phases $\alpha$ and $\beta$ is at $d = 0.25a$ (Fig. 2d).

To measure the fractional topological number at the edges, a finite 1D PC system is constructed. In the middle, PC1 has six unit-cells and is cladded by PC2 from left and right sides. In each side,

PC2 has two unit-cells (Fig. 2e). PC2 is kept in phase $\alpha$ with $d = 0.1a$, while for PC1 $d$ is tuned from $0.1a$ to $0.4a$ to achieve the topological transition from phase $\alpha$ to phase $\beta$. From finite-element simulations (see Methods), we find that when PC1 is in phase $\beta$, there are two edge states emerging in the photonic band gap, whereas there is no edge state if PC1 is in phase $\alpha$ (Figs. 2f-g) [23]. Calculated and measured wavefunctions for in gap edge states are given in Figs. 2h and 3b, respectively, showing excellent agreement with each other (except a shift in frequency due to the finite gap in the cladding [7]). Moreover, the bulk valence band have 10 eigenstates for the trivial case, as there are 10 unit-cells in the system. In comparison, there are 9 valence bulk states and 2 edge states for the topological case (consistent with the picture in Figs. 2d-e). If all the valence bulk states and edge states are filled, there will be (9+2)/2=5.5 modes in each half-sector, which is consistent with the fractional topological number $Q_e = \frac{1}{2}$.

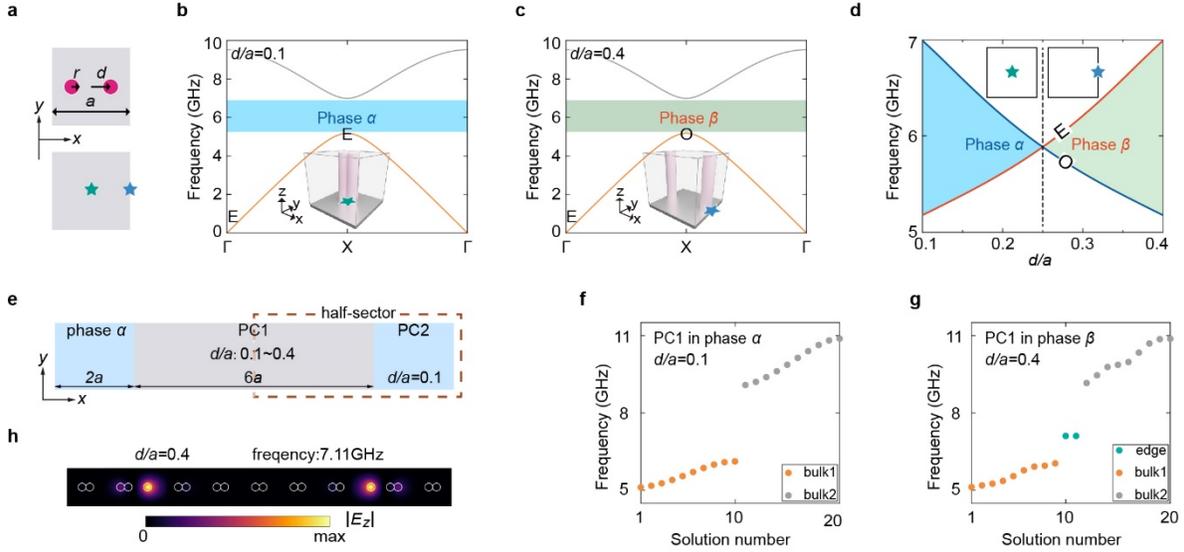

**Figure 2 | 1D PC, topological transition, and fractional topological number**. **a**, Upper panel: the unit-cell (gray region) of the 1D PC. Magenta dots denote the alumina pillars. Lower panel: Wannier centers in a unit-cell for the trivial (green star) and topological (blue star) cases. **b-c**, The first two photonic bands for the trivial (**b**, $d/a = 0.1$) and topological (**c**, $d/a = 0.4$) cases. O and E stand for the parity at the $\Gamma$ and X points, odd and even, respectively. Insets show the unit-cell structure and the Wannier center where the alumina pillars are painted as shallow-pink. **d**, Evolution of the photonic modes at the X point (phase diagram of the 1D PC). The red and blue curves represent the even- and odd-parity modes, respectively. **e**, Schematic of the experimental system with PC1 (gray region) cladded by PC2 (blue region). The dashed box indicates the half-sector where the LDOS and the mode charges are measured. **f-g**, Eigenstates spectrum

from finite-element simulations for two cases: (**f**) PC1 in phase $\alpha$ with $d/a = 0.1$, (**g**) PC1 in phase $\beta$ with $d/a = 0.4$. Orange (gray) dots denote the first (second) bulk band [denoted as bulk1 (bulk2)]. **h**, Calculated electric-field ($E_z$) pattern for edge state (green dots) in **g**.

Furthermore, when PC1 is in phase $\beta$, there is a fractional topological number $Q_e = \frac{1}{2}$ at each edge. To verify this, we measure the mode charge in each unit-cell for half of the system that contains an edge boundary (labeled as half-sector in Fig. 2e). The mode charge of the $j$-th unit-cell is defined as [7]

$$Q_j = \int_0^{f_{gap}} df \int_{j-th} d\boldsymbol{r}\ \rho(f,\boldsymbol{r}). \qquad (1)$$

Here, $f_{gap}$ is a frequency in the photonic band gap that is above the frequency of the edge states. The integration in real-space is within the $j$-th unit-cell. $\rho(f,\boldsymbol{r})$ is the photonic LDOS which depends on the frequency $f$ and the location $\boldsymbol{r}$. Here, the mode charge $Q_j$ characterizes how many photonic modes from the $j$-th unit-cell contribute to the valence band bulk states and the edge states (i.e., the photonic analog of the electronic charge when the valence band and the edge states are filled).

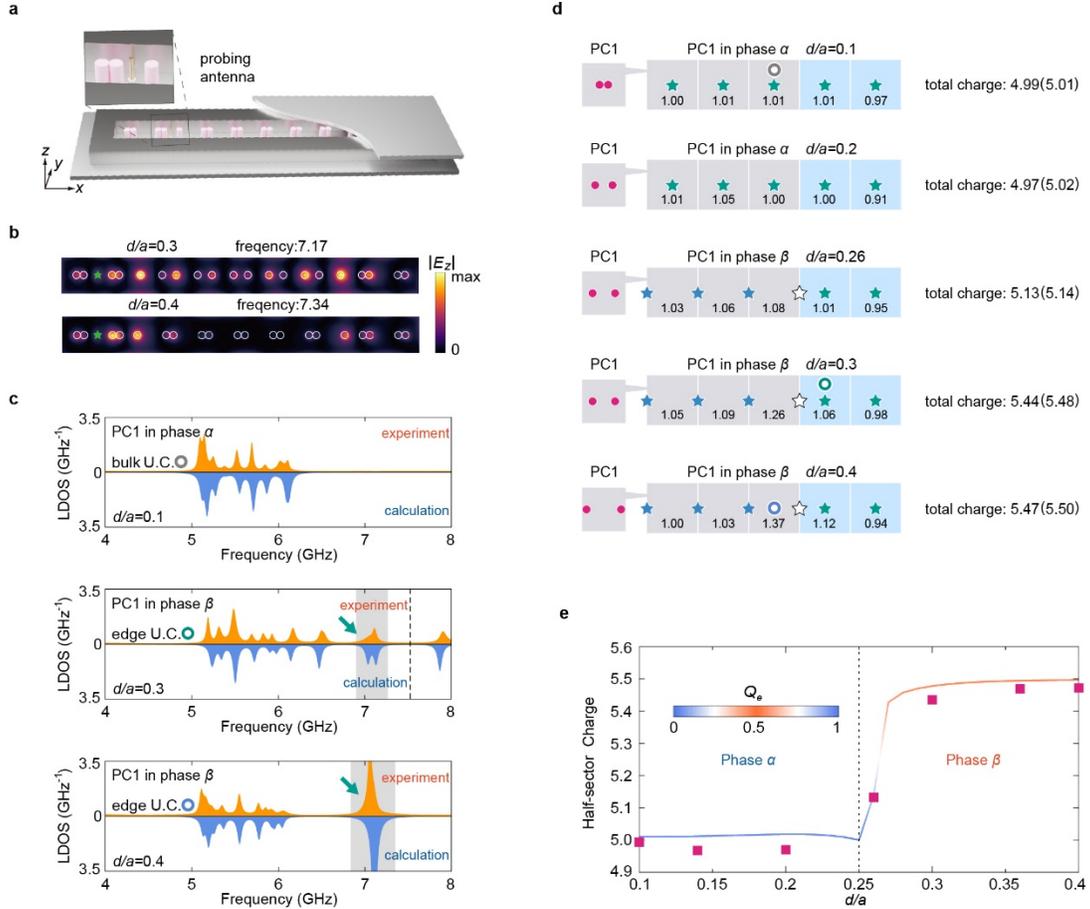

**Figure 3 | 1D PC, topological transition, and fractional topological number**. **a**, Experimental setup for the measurements of the photonic LDOS in 1D PC system. Zoom-in shows a photo with the probing antenna. **b**, Measured electric-field ($E_z$) profiles of edge states when PC1 is in phase $\beta$ with $d/a$ =0.3 and $d/a$ =0.4. Green stars denote the position where the excitation source is set. **c**, Measured photonic LDOS in three unit-cells (U.C.) in different configurations of which the positions are labeled in **d**. Orange (blue) shadows give the experiment (calculation) results. Green arrows indicate the edge states. **d**, Measured mode charges (given by the numbers in the unit-cells) in the right half of the system for five different configurations. The unit-cell structure of PC1 is illustrated in the leftmost. Green and blue stars give the Wannier centers for phase $\alpha$ and $\beta$, respectively. Hollow circles label the unit-cells of which the LDOS are presented in **c**. The measured total mode charge for each configuration is given in the right side along with the theoretical values (the numbers in brackets). **e**, Evolution of the total mode charge in the right half-sector with the distance $d$. The colored curve gives the calculation results (color represents the fractional topological number), while the squares give the experimental results.

The mode charge of each unit-cell is determined in both experiments and calculations based on Eq. (1) (see Methods). The photonic LDOS is measured through the classical analog of the Purcell effect which states that the radiative resistance of a subwavelength monopole antenna is proportional to the photonic LDOS [45]. By measuring the reflection spectroscopy $S_{11}$ of the antenna in the PC (see Fig. 3a) and in a benchmark 2D photonic system, we are able to determine quantitatively the photonic LDOS (see Methods). Figure 3c gives the photonic LDOS for three different unit-cells (labeled in Fig. 3d) as obtained from both experiments and calculations. For each unit-cell, the measured photonic LDOS agree reasonably with the calculated LDOS. The data also indicate that the LDOS at the edge boundary can be modified by tuning the topology of the PCs. As the system enters into the topological phase, the spectral weight in the valence band reduces, while the edge states emerge in the band gap. Due to their strongly localized nature, the edge states emerge as strong peaks in the LDOS curves in the band gap.

From the measured LDOS, we can obtain the mode charge of each unit-cell. The emergence of the fractional topological number is indicated by the fact that the total mode charge in half of the system has a fractional part. As shown in Figs. 3d-e, the total mode charge in the half-sector is close to 5.0 when $d < 0.25a$ (i.e., in phase $\alpha$). In the trivial phase, the mode charge of each unit-cell is approximately the number of bulk bands below the band gap, according to the Bloch's theorem. As the system crosses the topological transition at $d = 0.25a$, the total mode charge in the half-sector rises quickly from 5.0 to 5.5. At $d = 0.3a$, the half-sector already has a total mode charge of 5.44 (close to the calculated value of 5.48). For $d = 0.4a$, the half-sector has a total mode charge of 5.47 which is fairly close to the theoretical value of 5.5 in phase $\beta$. The consistency between the experiments, calculations, and theory indicates the experimental confirmation of the half quantum number $Q_e = \frac{1}{2}$ ---a well-known prediction of the Jackiw-Rebbi theory and the SSH model.

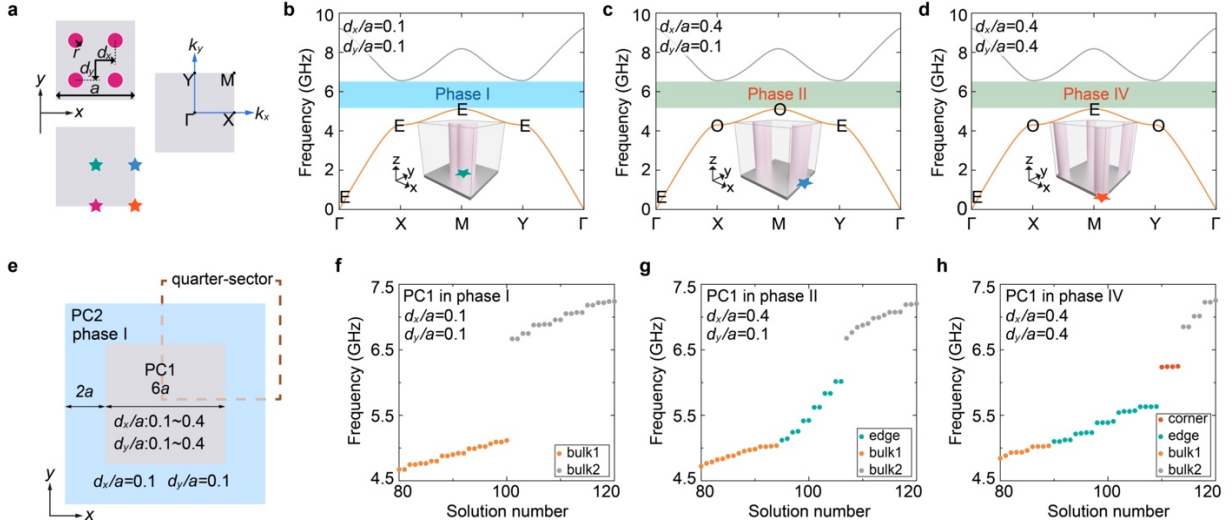

**Figure 4 | 2D PC, topological transition, and fractional topological number**. **a**, Upper-left panel: the unit-cell of the 2D PC. Magenta dots denote the alumina pillars. Lower-left panel: Wannier centers in a unit-cell for phase I (green star), II (blue star), III (magenta star), and IV (orange star). Right panel: Brillouin zone of the 2D PC. **b-d**, The first two photonic bands for phase I (b), II (c), and IV (d). O and E stand for the parity at the Γ, X, M and Y points, odd and even, respectively. Insets show the unit-cell structure and the Wannier center where the alumina pillars are painted as shallow-pink. **e**, Schematic of the experimental system with PC1 (gray region) cladded by PC2 (blue region). The dashed box indicates the quarter-sector where the LDOS and the mode charges are measured. **f-h**, Eigenstates spectrum from finite-element simulations for three cases: (**f**) PC1 in phase I, (**g**) PC1 in phase II, and (**h**) PC1 in phase IV. Orange (gray) dots denote the first (second) bulk band [denoted as bulk1 (bulk2)].

From the bottom panel of Fig. 3d, we find that the additional half mode charge is localized mainly at the edge boundary. Figure 3e summarizes the evolution of the fractional topological number across the topological transition with consistent results from experiments and calculations. This evolution confirms the fundamental picture in the SSH model: the half quantum number emerges at the topological boundaries. We emphasize that unlike the common conception that the fractional charges are associated with edge or corner modes, the truth is that these boundary modes in fact do not carry fractional charges [23]. Moreover, it is worth clarifying that these boundary modes need both the spatial symmetry (such as inversion and rotation symmetries) and the particle-hole symmetry to protect them. In contrast, the fractional charges only need the spatial symmetry to protect them---they are thus more robust as shown in recent experiments [5-8, 46, 47]. Therefore, although disorders can create trivial defect modes or move the edge or corner modes into bulk continua, they do not affect the fractional charges as long as the spatial symmetry is preserved---which is absolutely necessary to define the symmetry protected topology. These understandings have been established in recent theoretical works [23-29]. We also show in Supplementary Note 3 via numerical simulations that defect modes induced by disorders cannot lead to half quantum numbers. Finally, the continuous change of the fractional topological number across the topological transition is due to the finite size of the system. Such change could be abrupt in infinite systems.

We clarify several points here. First, there seems to be a gap between the continuous model in the Jackiw-Rebbi theory and the discrete SSH model, despite that they give the same predictions. This gap can be filled by considering the adiabatic continuity when tuning the system: One can gradually tune the system towards a small band gap without changing its symmetry and topological invariants (and hence the topological properties). In the small band gap limit, the band gap can be described approximately by the massive Dirac equations as in the Jackiw-Rebbi model which then unifies the Jackiw-Rebbi theory and the discrete SSH model. Second, in this work we also use the Wannier center picture, which is well-defined for periodic systems. Certainly, when one considers finite-sized systems, the Wannier center picture is not precise. Nevertheless, we can invoke again the adiabatic continuity to solve this issue: Theoretically, one can gradually switch off the weak intra-unit-cell couplings without changing the symmetry and topological properties of the system. In such a limit, one recovers the local orbital picture provided by the Wannier centers. Within this picture, the fractional charges and topological boundary states emerge naturally, as shown in Ref. [23]. Therefore, through the adiabatic continuity, we can put these pictures on equal ground and use them to analyze our systems.

We now study the fractional topological number in 2D PCs. The 2D SSH model is realized via a configurable PC with four pillars in each unit-cell which are located at $(\pm d_x, \pm d_y)$ (the origin of the coordinate is at the unit-cell center). The PC has mirror symmetries with respect to the *x* and *y* directions and the inversion symmetry. These symmetries protect the topology of the photonic bands and restrict the allowed positions of the Wannier center (see Supplementary Note 2). The band topology can be tuned by the two parameters $d_x$ and $d_y$, as shown in Figs. 4b-4d. Since phase

III and phase II can be transformed into one another via 90° rotation, we skip phase III in our investigations. The results in Figs. 4b-4d indicate that all phases I-IV in the 2D SSH model can be realized in the 2D configurable PC [40-42].

To observe the fractional topological numbers in these phases, we construct a finite 2D PC as schematically shown in Fig. 5a. PC1 in the center with 6 × 6 unit-cells is cladded in all directions by PC2 of the thickness $2a$. While PC2 is kept in phase I with $d_x = d_y = 0.1a$, PC1 is tuned to experience phases I, II and IV. The calculated photonic spectra for the three cases are shown in Figs. 4f-h which manifest the following features (see more details in Supplementary Note 4): When PC1 is in phase I (the trivial phase), there is no topological boundary state (Fig. 4f). In contrast, when PC2 is in phase II, topological edge states emerge in the photonic band gap (Fig. 4g). As PC2 enters into phase IV, topological edge and corner states emerge simultaneously (Fig. 4h). In addition, when PC2 is in phase I, there are 100 valence bulk states. As PC2 enters into phase II, there are 94 valence bulk states and 12 edge states, which is consistent with the filling anomaly with 12 edge unit-cells. If all the valence bulk states and edge states are filled, there will be (94+12)/4=26.5 modes per quarter of the system, which is consistent with the edge fractional topological number $Q_e = \frac{1}{2}$ (There are 3 edge unit-cells per quarter sector, therefore there are 25+3/2=26.5 modes in each quarter-sector). When PC2 is in phase IV, there are 89 valence bulk states, 20 edge states and 4 corner states, which agrees with the filling anomaly with 20 edge unit-cells and 4 corner unit-cells. If all the valence bulk and edge states are filled, there will be (89+20)/4=27.25 modes per quarter of the system, which gives the corner fractional topological number of $Q_c = \frac{1}{4}$ (See Supplementary Note 5 for details).

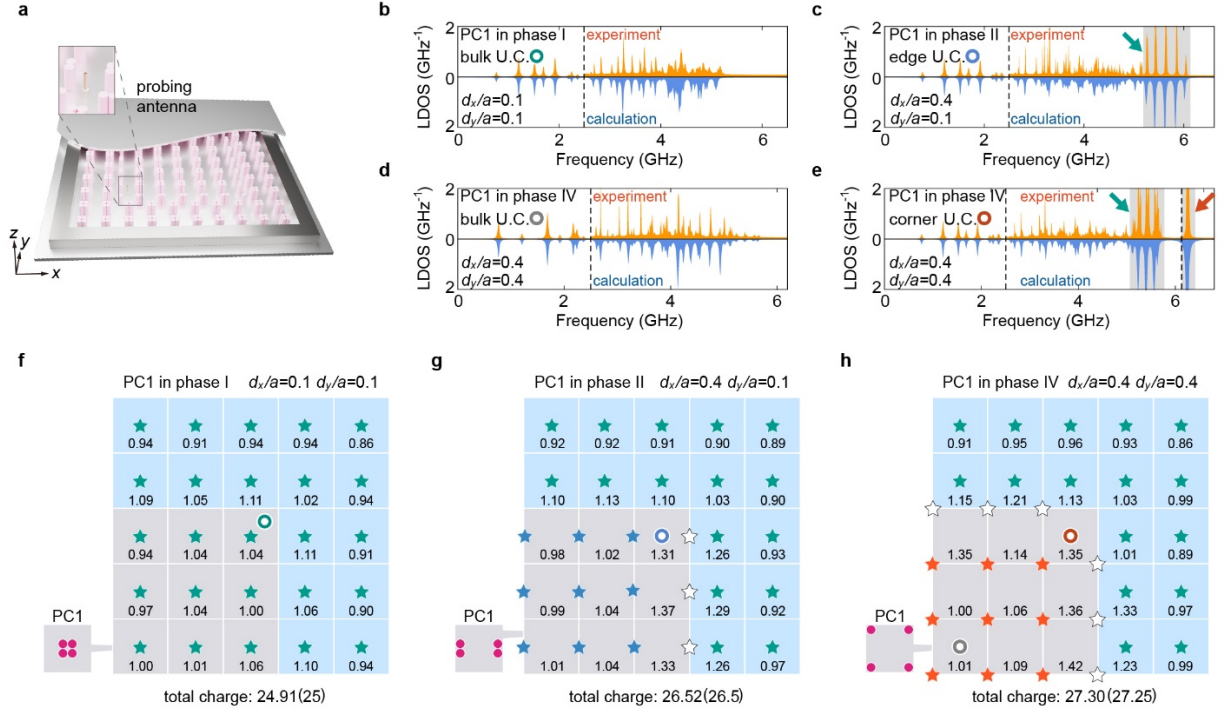

**Figure 5 | 2D PC, topological transition, and fractional topological number**. **a**, Experimental setup for the measurements of the photonic LDOS in 2D PCs. Zoom-in shows a photo with the probing antenna. **b-e**, Measured photonic LDOS in four unit-cells (U.C.) in different configurations of which the positions are labeled in **f-h**. Orange (blue) shadows give the experiment (calculation) results. Green (red) arrows indicate the edge (corner) states. **f-h**, Measured mode charges (given by the numbers in the unit-cells) in the upper-right quarter of the system for three different configurations. Unit-cell structure of PC1 is illustrated in the lower-left. Stars give the Wannier centers. Hollow circles label the unit-cells of which the LDOS are presented in **b-e**. The measured total mode charge for each configuration is given at the bottom along with the theoretical values (the numbers in brackets).

The experimental setup for the 2D PC system is shown in Fig. 5a. A probing antenna is used to detect the photonic LDOS from which the mode charge and fractional topological numbers can be determined. Four examples of the measured LDOS are shown in Figs. 5b-e which manifest dramatical changes in photonic LDOS by tuning the band topology via the geometric parameters $d_x$ and $d_y$. In all these examples, the measured LDOS agree well with the calculated LDOS. Due to the limitation of our measurement system, we can only cover frequencies higher than 2.5 GHz (labeled by the dashed lines) with both the valence band and the edge states. Although there are sharp peaks in the figures, we collected about ten thousand data points for each curve to resolve all peaks. These peaks mostly agree with the peaks from calculations. The measured LDOS for other unit-cells are presented in Supplementary Note 6 which all show reasonable agreement with the calculations (see Supplementary Note 7 for methods and details). Figures 5b-e show the decrease of the spectral weight of the valence band and the emergence of the topological edge and corner states in the band gap. In Fig. 5c, the nearly equally separated peaks in the band gap are consistent with the edge states spectrum in Fig. 4g. The broadened peaks in the band gap in Fig. 5e are consistent with the increase of the number of edge states and corner states when PC1 enters into phase IV. Meanwhile, the LDOS in the bulk unit-cells also exhibit tails in the band gap due to the spatial extension of the wavefunctions of the edge and corner states, as shown in Fig. 5d.

From the measured LDOS, we obtain the mode charge distributions in a quarter-sector of the PC system for the three cases (PC1 in phases I, II, and IV, respectively; see Supplementary Note 8 for details) and present the results in Figs. 5f-h. Figure 5f indicates that when PC1 is phase I, the measured total mode charge in the quarter sector is 24.91 which is close to the theoretical value of 25. When PC1 is in phase II, the measured total mode charge is 26.52 which agrees with the theoretical value of 26.5 (Fig. 5g). Finally, when PC1 is in phase IV, the measured total mode charge in the quarter sector is 27.30 which is near the theoretical value of 27.25 (Fig. 5h).

## Summary and discussions

Using tunable dielectric PCs, we reveal the fractional topological numbers at edges and corners as the topological observables in photonics. In particular, the long-sought half quantum number in the well-known 1D SSH model is confirmed in our photonic experiments. In contrast to previous studies focusing on topological edge and corner states, here, the fractional topological numbers depending solely on the bulk states are the key features. Our study connects the photonic LDOS, an important property of engineered electromagnetic environments, to the fractional topological numbers that characterize the photonic band topology. This connection also unveils topological mechanisms for the manipulation of photonic LDOS which are valuable for studying topological effects in optoelectronics, lasing, and quantum electrodynamics.

## Materials and methods

Numerical simulations of photonic eigenstates and photonic band gaps are carried out using the radio-frequency module in the commercial finite-element software COMSOL MULTIPHYSICS. The simulations are performed under transverse-magnetic harmonic modes (i.e., the electric-field is always perpendicular to the *x-y* plane, along the *z* direction). In band structure calculation, the relative permittivity of cylinders is set as $\varepsilon_r$=8.4.

All the samples in experiment are fabricated by pasting low-loss dielectric cylinders ($\varepsilon_r$=8.4-0.02i) which is constructed of Aluminium oxide ($Al_2O_3$) doped with Chromium oxide ($Cr_2O_3$) on pre-printed Polyvinyl Chloride substrates. The cylinders are machined into a size of $\pi \times 2^2 \times 10$ mm$^3$ with a tolerance of +0.01mm. The 1D samples have a sandwich structure of PC2-PC1-PC2 with a finite size where cell's number are 2, 6 and 2, respectively. The 2D samples are fabricated into a size of 6×6 cells PC2 with PC1 coating around it, which eventually constructs a 10×10 cells symmetric structure. In the experimental observation, the fractional mode charge is calculated from the LDOS which is deduced from the measured reflection coefficient (see Supplementary Note 8). The experimental setup mainly includes a vector network analyzer (Agilent PNA-X N5247A) and a paralleled plate waveguide of 300mm×600mm×10.2mm along the *x*, *y,* and *z* directions, separately, to simulate a quisi-2D photonic environment. During the measurements, the sample surrounded by aluminum frame is placed in the paralleled plate waveguide. The restriction of the frame does help for photons leakage prevention. The electromagnetic wave's feeding and detection are both accomplished by a short monopole antenna mounted on a sub-miniature A connector.

## Acknowledgements

Y.P., F.-F.L., C.L. and S.L. thank the National Natural Science Foundation of China (NSFC) (Grant Nos. 62171215 and 62001212), Natural Science Foundation of Jiangsu Province (BK20201249), the Project of Young Scientific and Technological Talents Promotion and the Projects Funded by the Priority Academic Program Development of Jiangsu Higher Education Institutions and Jiangsu Provincial Key Laboratory of Advanced Manipulating Technique of EM waves. Y.L. and J.-H.J. are supported by the Jiangsu Province Specially-Appointed Professor Funding, the National Natural Science Foundation of China (Grant nos. 12125504 and 12074281) and the Project Funded by the Priority Academic Program Development of Jiangsu Higher Education Institutions.

## Conflict of interests

The authors declare no conflict of interests.

## Contributions

J.-H.J. and Y.P. initiated the project and guided the research. J.-H.J. and Y.L. established the theory. Y.L. performed the numerical calculations and simulations. C.L., F.-F.L., S.L., J.-H.J. and Y.P. designed and achieved the experimental measurements. All the authors contributed to the discussions of the results and the manuscript preparation. J.-H.J., Y.P. and Y.L. wrote the manuscript and the Supplementary Information.